\documentclass[aps,prb,twocolumn,showpacs,superscriptaddress]{revtex4}

\usepackage{graphicx} 
\usepackage{multirow}
\usepackage[T1]{fontenc}
\usepackage[latin9]{inputenc}
\usepackage{amsmath}
\usepackage{amssymb}
\usepackage{color}

\date{\today}
\begin{document}
\title{Graphene Nanomeshes: Existence of Defect-Induced Dirac Fermions on Graphene Host Matrix}
\author{H. \c{S}ahin}
\affiliation{UNAM-Institute of Materials Science and
Nanotechnology, Bilkent University, 06800 Ankara, Turkey}
\author{S. Ciraci}\email{ciraci@fen.bilkent.edu.tr}
\affiliation{UNAM-Institute of Materials Science and
Nanotechnology, Bilkent University, 06800 Ankara, Turkey}
\affiliation{Department of Physics, Bilkent University,
06800 Ankara, Turkey}

\date{\today}


\begin{abstract}
Motivated by the state of the art method for fabricating high
density periodic nanoscale defects in graphene, the structural,
mechanical and electronic properties of defect-patterned graphene
nanomeshes including diverse morphologies of adatoms and
holes are investigated by means of first-principles calculations
within density functional theory. It is found that various patterns
of adatom groups yield metallic or semimetallic, even semiconducting
behavior and specific patterns can be in a magnetic state. Even though
the patterns of single adatoms dramatically alter the electronic structure
of graphene, adatom groups of specific symmetry can maintain the Dirac
fermion behavior. Nanoholes forming nanomesh are also investigated.
Depending on the interplay between the repeat periodicity and the geometry
of the hole, the nanomesh can be in different states ranging from metallic
to semiconducting including semimetallic state with the bands crossing
linearly at the Fermi level. We showed that forming periodically repeating
superstructures in graphene matrix can develop a promising technique to
engineer nanomaterials with desired electronic and magnetic properties.
\end{abstract}

\maketitle

\section{Introduction}

Propagation of electron waves through the honeycomb lattice
attributes exceptional features to graphene.\cite{geim}
Conduction of electrons within one-atom-thick layer with
minute scattering makes the observation of quantum effects
possible even at room temperature.\cite{dai-transistor,berger}
Experimental investigations have reported the
observation of half-integer quantum Hall effect
for carriers in graphene and possible magneto-electronic
device applications.\cite{zhang} Most of the unique
properties of graphene are related to its monolayer lattice
structure, linearly crossed $\pi$ bands at Fermi level
with electron-hole symmetry. Recently, we showed that
the honeycomb structure with linear band crossings at
Dirac points is also common to Si and Ge.\cite{Silicene, hasan-ansiklopedi}

In an effort to make semimetallic graphene suitable for electronic
applications, it has been functionalized to generate band gaps.
It was theoretically shown that it is possible to induce band-gap
opening produced by the adsorption of atomic hydrogen on
graphene by choosing specific adsorption periodicity.
\cite{castra}  It was also experimentally shown that
B- and N-doped graphenes can be synthesized to
exhibit p- and n-type semiconducting properties
that can be systematically tuned with the dopant
concentration.\cite{pancha} Effect of hydrogenation
and the transition metal atom adsorption on the
transport properties of graphene was also investigated
theoretically.\cite{hasan-prb2008, hasan-jap2010} Using
symmetry arguments and tight binding calculations it was
shown that the periodic structure of defects (such as B and N
impurities) on graphene can exhibit semimetallic and
semiconductor behavior.\cite{tantar} Moreover, weak
perturbation potential forming a large hexagonal lattice in a
two dimensional electron gas was shown to lead a massless
Dirac fermion Hamiltonian with linearly crossing bands at
Dirac points.\cite{stroud, park,gibertini}

A majority of the current studies on graphene is devoted to
its chemical modification to create derivatives with
different structures and properties. So far three known
derivatives of graphene have been successfully achieved in
chemical reactions: graphene oxide (GO),
\cite{dikin,stankovic1,eda,navarro} graphane
(CH)\cite{elias,sofo,hasan-apl,flores} and recently
fluorographene (CF).\cite{nair, cheng, hasmeh} Although GO is
a wide band gap material that is important for device applications,
its atomic structure, wherein the carbon atoms are decorated with
epoxides, alcohols and carboxylic acid groups, is not
suitable for nanoscale manipulations. CH obtained by
exposing carbon honeycomb structure to hydrogen plasma, is
another example of the graphene-based chemical derivative.
Upon the hydrogenation, semimetallic graphene is converted
into an insulator. CF, the two dimensional counterpart of teflon,
is the most recent focus of graphene research.

Much recently, the fabrication of large graphene sheets
having high-density array of nanoscale holes, called graphene
nanomeshes (GNMs),\cite{bai} has been the landmark in
controlling the electronic properties at nanoscale.
Additionally, the formation of one-dimensional periodic
Stone-Wales type defects producing metallic nanowires on
graphene matrix has also been reported.\cite{batzill} These
recent advances have made mesh configuration a controllable
parameters to monitor physical properties of nanostructures.
\cite{bai, batzill, balog} Earlier, interesting effects of periodically
repeating holes in the electronic and mechanical properties of
graphene nanoribbons were predicted from the first-principles
calculations.\cite{topsakal}

In this paper, we apply supercell method to reveal the electronic, magnetic
and mechanical properties of graphene which is patterned by various adsorbates 
or holes. The atomic structure of all adsorbates and holes are obtained after 
extensive structure optimization. These periodically repeating superstructures or
nanomeshes display properties which are rather different from those of
graphene. We showed that not all patterns of adsorbates or hole with a
2D hexagonal lattice on graphene have linear band crossing, but only those
which have a specific rotation symmetries. However, depending on the size of
patterns or holes and the repeat periodicity, a GNM can be in different
states ranging from to semiconducting including semimetallic with linear
band crossing at the Fermi level. 

\section{Computational Methodology}

The present study revealed crucial effects of the point group symmetry of
nanomesh on the resulting properties. Here, we start with a brief discussion
of hexagonal symmetry and apply simple tight binding model of $\pi$-orbitals
to reveal the effect of lattice symmetry on the band crossing.\cite{tb2,tb1}
Graphene has the space group P6/mmm and point group symmetry D$_{6h}$. At
the-$\Gamma$ point, the group of the wave vector is isomorphic to the point
group D$_{6h}$.\cite{dressel} However, irreducible representation of the
wave vector point group turns into D$_{2h}$ and D$_{3h}$ at high symmetry
points $M$ and $K$ (or $K^{'}$), respectively. It was shown that the tight
binding Hamiltonian with nearest neighbor hopping parameter, $t=2.7$ eV

\begin{figure}
\includegraphics[width=8.5cm]{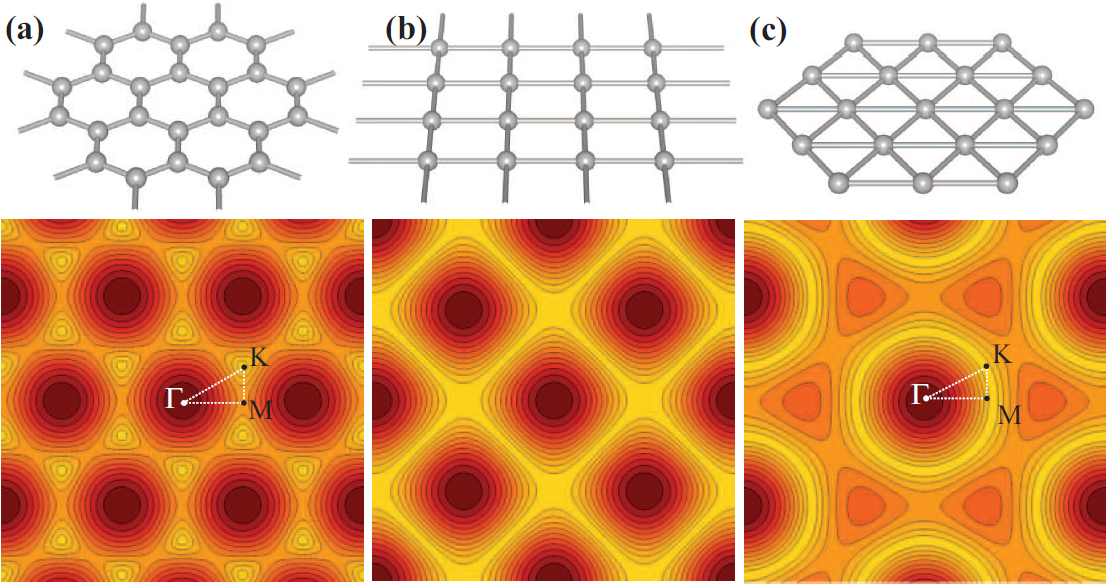}
\caption{Contour plots for band gap. (a) Perfect graphene.
(b) A crystal having square lattice with single atom in the unit cell.
(c) A crystal having hexagonal lattice with a single atom in the unit cell.
Band crossing occurs along the yellow/light contours on which the
band gap becomes zero. Band gap takes its maximum value at brown/dark contours.}
\label{tb}
\end{figure}

\begin{footnotesize}
\begin{equation}
 H=\sum_{i}^{}\epsilon_{i}c_{i}^{\dag}c_{i}+ t \sum_{i,j}^{}(c_{i}^{\dag}c_{j}+H.c).
\end{equation}
\end{footnotesize}

well approximates the $\pi$-bands of perfect graphene.\cite{tb2, tb1}
Here $c_{i}^{\dag}$ ($c_{i}$) is the creation (annihilation) operator of a
$\pi$ electron at the lattice site \textit{i}. The first term is the on-site
energy of each carbon atom and equals to energy of the $2p_{z}$ orbital.
Energy eigenvalues of graphene and other two hypothetical crystal having
square and hexagonal lattices with single atom in the cell are calculated
and the contour plots their energy band gap in BZ are shown in Fig.\ref{tb}.
For graphene, energy dispersion is linear at the vicinity of the $K$-
symmetry (Dirac) points and the Fermi velocity, which is linearly dependent
to nearest neighbor interaction parameter, can be given by the expression
$v_{F}=3td/2\hbar$. \cite{revmod}

Hypothetical square lattice has semimetallic band structure as in Fig.
\ref{tb}(b). In the reciprocal space the band gap is closed along the
boundaries of squares. However, differing from graphene, these bands do not
have linear dispersion. The structure having hexagonal lattice with single
$\pi$-orbital per unitcell is shown in Fig.\ref{tb}(c). Such a structure
having a single $\pi$-orbital in each unitcell may also be realized by
adatom saturation of one type (A- or B-type) carbon atoms of graphene. In
this case, structure has three-fold rotation symmetry and hence six nearest
neighbors. In this case, while saturation yields a dramatic change in the
electronic structure, linear band crossings at Fermi level still occur at
the points on a circle of radius $R=\frac{4\pi}{3\sqrt{3}}$ in BZ. As shown
in Fig. \ref{tb}(c), this circle passes through the corners of hexagonal BZ of graphene. The Fermi velocity is calculated to be in the order of 10$^{6}$
$m/s$ in the vicinity of band crossing points.

While the tight binding model allowed us to understand the general features
of the band gap in different lattices, it fails to account for the
reconstruction and rebonding near the defect. In the rest of the paper
we perform calculations from the first-principles to investigate various
types of defects. To this end we carried out spin-polarized plane wave
calculations\cite{vasp1,vasp2} using local density  approximation
(LDA)\cite{lda} and projector augmented wave (PAW)\cite{paw} potential.
Patterns of defects are treated using supercell geometry, where a minimum of
10 \AA~ vacuum spacing is kept between the adjacent graphene layers. Kinetic
energy cutoff, Brillouin zone (BZ) sampling are determined after extensive
convergence analysis. For the plane-wave basis set, the kinetic energy
cutoff is taken to be $ \hbar^2 | \mathbf{k}+\mathbf{G}|^2 / 2m = 500$ eV.
For partial occupancies Methfessel-Paxton smearing method\cite{methfessel}
is used. The convergence criterion of self-consistent field calculations is
$10^{-5}$ eV for total energy values. By using the conjugate gradient
method, all atomic positions and the size of unitcell were optimized until
the atomic forces were less than 0.05 eV/\AA. Pressures on the lattice unit
cell are decreased to values less than 1 kB.

\begin{figure*}
\includegraphics[width=17.9cm]{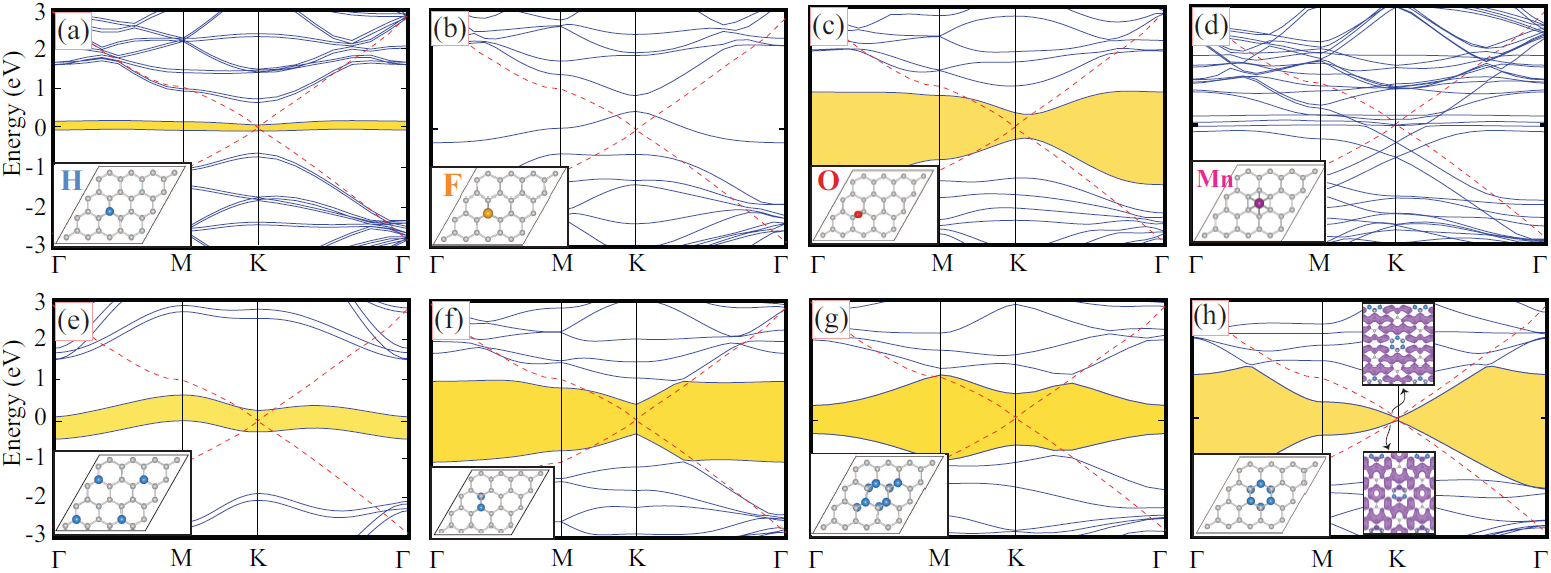}
\caption{Spin-polarized energy band structure of a
periodic patterns consisting of the (4x4) supercells each
having a single (a) hydrogen; (b) fluorine; (c) oxygen; (d)
manganese adatom. (e) A similar hydrogen pattern forming the
(2x2) supercell on graphene host matrix allowing significant
coupling between adatoms. (f-h) Periodic patterns of two,
eight and six hydrogen atoms in the (4x4) graphene supercell,
respectively. Isosurface of charge density of bands crossing
near the $K$-point are shown by insets. For the sake of
comparison, linearly crossing
$\pi$ and $\pi^{*}$-bands of perfect graphene host matrix are
also superimposed in the band structures. The zero of energy
is set at the Fermi level $E_F$. The band gaps are shaded in
yellow. All the bands presented in BZ corresponding to the
(4x4) supercell.}
\label{f2}
\end{figure*}

\section{Adatom Patterned Graphene Nanomeshes}

In this section we show how a periodic decoration of graphene by adatoms
modifies the electronic structure. Here we considered H, F, O and Mn
adatoms, which are adsorbed at different sites and form a (4x4) supercells
on graphene host matrix as shown in Fig.\ref{f2}(a)-(d). We did not consider
the interstitial or substitutional decoration, since experimental studies
treating various foreign atoms, such as N\cite{pancha}, O\cite{dikin},
H\cite{elias}, F\cite{nair}, C\cite{ad-C}, Co\cite{ad-Co}, Fe and
Gd\cite{ad-Fe-Gd}, Au and Pt\cite{ad-Au-Pt}, revealed that these atoms
prefer to be adsorbed at various sites on the surface of graphene, but none
of them is adsorbed at the interstitial sites, nor is substituted for carbon atom.

Owing to the unpaired electron, a single hydrogen adatom adsorbed to the
(4x4) supercell has a spin polarized, semiconducting ground state with a net
magnetic moment of $\mu$=1 $\mu_B$. Upon the adsorption of a hydrogen atom
band structure of graphene changes dramatically. Instead of linear crossing
of $\pi$ and $\pi^{*}$-bands at Fermi level, dispersionless impurity bands
occurs with 0.1 eV indirect band gap. Similar to hydrogen atom, the most
favorable adsorption site for a F atom on graphene is the top site of carbon
atoms. Upon the adsorption of a F atom, $sp^{2}$ bondings of three C-C bonds
below F atom are dehybridized and form tetrahedrally coordinated four
$sp^{3}$ type bondings. Differing from the decoration of H adatom, ground
state is nonmagnetic and as a result of odd number of electrons F decorated
graphene becomes metalized as shown in Fig. \ref{f2}(b). The band crossing
at the K-points does not occur, since F adsorbed at the top of C atom
changes the sixfold rotation symmetry to the threefold rotation symmetry.

Oxygen atom favors the bridge site between two underlying C atoms. Upon the
adsorption of an oxygen at the bridge site two underlying C atoms become
buckled by 0.36 \AA. C-C and C-O bonds are calculated to be 1.51 and 1.44
\AA, respectively. Resulting structure is a nonmagnetic semiconductor with
0.63 eV direct band gap (Fig. \ref{f2}(c)). Valence and conduction band
edges occurs between the $K$- and $\Gamma$-points. The adatom at the bridge
site breaks the six fold rotation symmetry and hence hinders the linear band
crossing.

The situation is different in the case of Mn, which is adsorbed above 
the center of a hexagon in graphene matrix and induces negligible deformation. 
Only six nearest C atoms raise slightly to higher (0.02 \AA) position 
relative to the plane. Localized, non-bonding Mn-$3d$ orbitals form flat bands 
near the Fermi level. On the other hand, the sixfold rotation symmetry is maintained 
even after Mn atom adsorbed at the hollow site above the center of hexagon. 
Accordingly, the metallic structure with a net magnetic moment of 3 $\mu_{B}$ 
per cell allows linear crossing of graphene bands at the $K$-points in Fig. \ref{f2}(d).

Having discussed the effect of periodic decoration by single
adatoms, we next consider the periodic patterns of adatom
groups. In Fig. \ref{f2}(e) we show electronic band structure
corresponding to relatively denser hydrogen coverage
(C:H$=$8). Such a nanomesh created by one-sided decoration of
four hydrogen atoms in the (4x4) supercell gives rise to a
relatively dispersive bands and a net magnetic moment of 4
$\mu_B$ per supercell. Since the six fold rotation symmetry of
graphene is broken by adsorbed H atoms, linear crossing of
bands do not occur. In a decoration involving two sides of
graphene, where two adjacent C atoms of different sublattices
are saturated from different sides as shown in Fig. \ref{f2}
(f). Since an equal number of A- and B-sublattice atoms are
saturated, the structure is nonmagnetic semiconductor with a
band gap of 0.8 eV. Another pattern  derived from a graphane
like domain consisting of 8 H atoms in  Fig. \ref{f2}(g)
results in a band gap of 0.7 eV at $\Gamma$-point, but larger
gap of 1 eV at the $K$-point. This nanomesh presents an electronic
structure rather different both from graphene and graphane.
The electronic structure is, however, different for
a pattern of six H atoms, which saturate six carbon atoms at
the corner a hexagon alternatingly from different sites;
namely three of them adsorbed to A-sublattice from one side,
remaining three adsorbed to B-sublattice from the other side.
Even though sixfold rotation symmetry has changed to $S_6$ symmetry,
both point group symmetries allow linear band crossing
as seen in Fig. \ref{f2}(h). This case demonstrates the
crucial role played by the intrinsic symmetry of the pattern
in determining the electronic structure.
\cite{park,tantar,stroud}

\begin{figure}
\includegraphics[width=8.5cm]{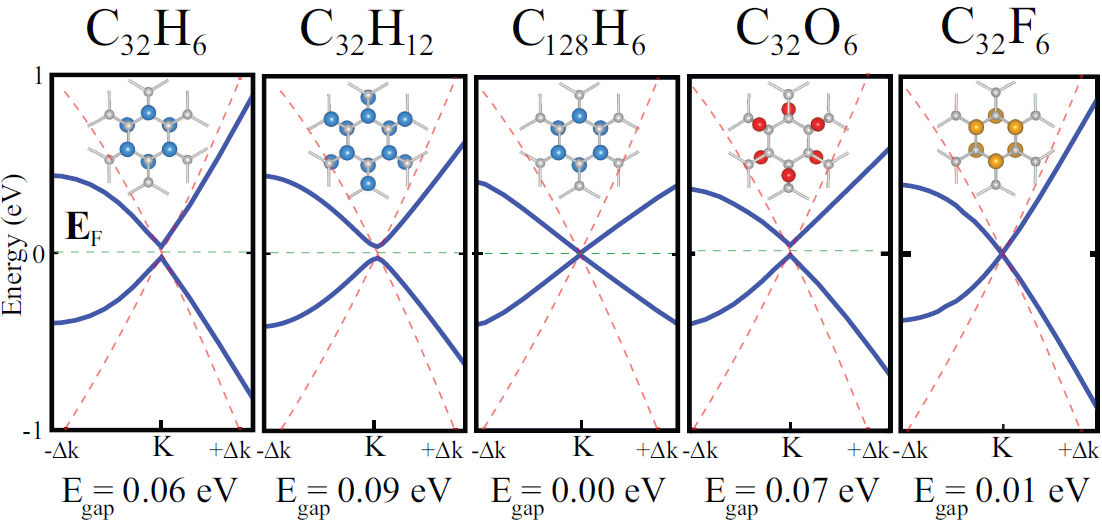}
\caption{Band structures showing the effects of the coupling
between various patterns as a function of their size and the
size of mesh (supercell). The linearly crossing of $\pi$- and
$\pi^{*}$-bands of graphene are shown by red/dashed lines.}
\label{f3}
\end{figure}

Let us now take a closer look at the triangular patterns of adatom groups
that have hexagonal symmetry. In Fig. \ref{f3}, we plot these $\pi$- and
$\pi^{*}$-bands in the vicinity ($\Delta$k=0.03 \AA$^{-1}$) of $K$-symmetry
point for different patterns on supercells of different sizes; namely
the (4x4) and (8x8) (C$_{128}$ and C$_{32}$) supercells. H$_{6}$ and
H$_{12}$ patterns which have 0.06 and 0.09 eV band gap at $K$-point,
indicates that the band gap opening increases with increasing pattern size
and hence increasing coupling. In the (8x8) supercell the interaction
between periodically repeating $H_{6}$ patterns is hindered and hence
linear crossing of $\pi$- and $\pi^{*}$-bands at $K$-symmetry point similar
to bare graphene is attained. By expressing these bands as
$\bf{k}=\bf{K}+\bf{q}$ and by neglecting the second order terms with respect
to $q^{2}$, the dispersion of the energy bands can be given as $E(\bf{q})
\simeq v_{F}\hbar|q|$. Here the Fermi velocity $v_{F}$ is calculated as
0.7x$10^{6}$ m/s (i.e. $\sim0.6$ of the Fermi velocity calculated for Dirac
Fermions in perfect graphene).

We also present an example of O$_{6}$ pattern analogous to
H$_{12}$ decoration of (4x4) supercell (C$_{32}$H$_{12}$) of
graphene in Fig. \ref{f3}. In this pattern three O atoms are
bound alternatingly to bridge site at one site, the remaining
three to other site. Here it is seen that denser O patterns
are not favored due to strong O-O repulsion. Linear crossing
of bands at $K$-point can also be achieved by O adatoms
forming a periodically repeating O$_6$ pattern. Small band gap
in Fig. \ref{f3} can be closed if the supercell size is
increased to hinder coupling between them.  This example
implies that the patterns similar to that done by using H
atoms can be created by the passivation of $p$-orbital
electrons with O atoms. In the case of fluorination of (4x4)
graphene by F$_{6}$ decoration, similar to H$_6$, linear $\pi$
and $\pi^{*}$-bands gets closer with very small (0.01 eV) band
gap at $K$-point. Finally, the isosurface charge densities of
these linearly crossing bands near $K$-point indicates that
they mainly originate from graphene $\pi$-orbitals with small
mixing from the adatom (see Fig. \ref{f2} (h)).

\section{Hole Patterned Graphene Nanomeshes}

\begin{figure}
\includegraphics[width=8.5cm]{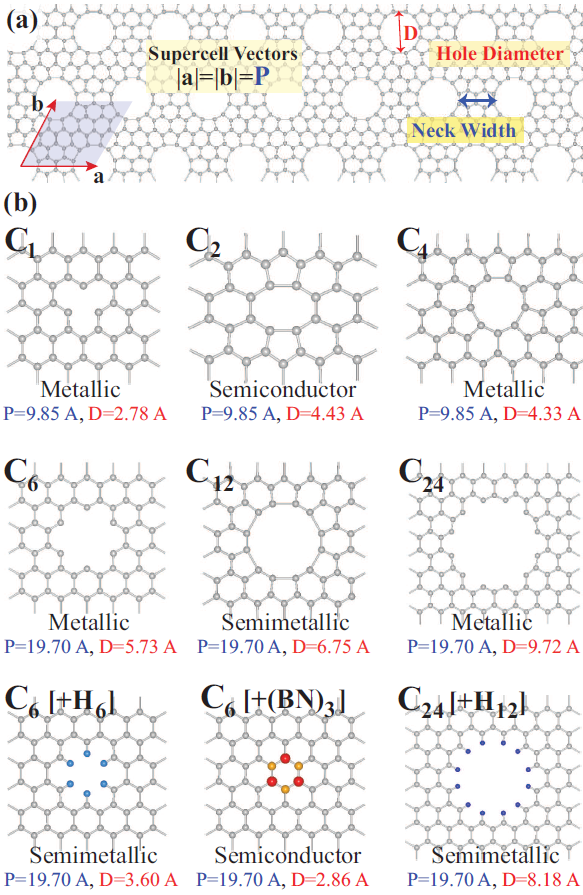}
\caption{(a) Structural parameters for a nanomesh of C$_{n}$
hole. (b) Large supercells of hexagonal lattice  each
containing single hole of C$_{1}$, C$_{2}$, C$_{4}$, C$_{6}$,
C$_{12}$, C$_{24}$. The nanomeshes in the third row  are
obtained by saturating C$_{6}$ and C$_{24}$ holes by hydrogen
and also by B and N atoms alternatingly. }
\label{f4}
\end{figure}

Similar to adatom patterns on graphene, nanomeshes generated
from the holes periodically patterned on graphene matrix
exhibit also interesting features. This conclusion drawn from
theoretical calculations are in line with the findings
obtained from the fabrication of GNMs by means of block-copolymer
assisted nanopatterning process. It has been shown
that GNMs having high-density periodic array of holes display
promising advantages relative to existing graphene
devices.\cite{bai, batzill} Thanks to the advances in the preparation
of high quality nanoscale hard masks\cite{bai, batzill, jacs}
in laboratory conditions, theoretical studies in this
field become more relevant for applications. Here we carry
out calculations for the holes having 1-2 nm repeat period
and 2-10 \AA~ diameter. In Fig. \ref{f4}(a) we describe the
geometric parameters of C$_{n}$ hole defects forming a
hexagonal lattice, where $n$ denotes number of C atoms
removed from graphene matrix to make a hole. For nanoholes,
we define the hole size as the maximum diameter of the C$_n$
hole defected region. After the creation of a C$_{1}$ defect
(i.e. single C vacancy), Jahn-Teller type distortion changes
the positions of surrounding C atoms slightly. The resulting
structure attains a net magnetic moment of 1 $\mu_{B}$. The origin 
of magnetism in defected graphene sheets and the character of 
electronic states induced by the vacancy resulting in flat bands 
around Fermi level have been investigated by some 
recent studies.\cite{Lieb, yazyev, nanda} Upon the removal 
of C atoms ($n$>1) graphene is reconstructed to
result in a significant modification in atomic configuration
around the hole. For example, after the relaxation of atomic
structure, C$_{2}$ defected region becomes an octagon-shaped
hole surrounded by 2 pentagons and 6 hexagons.  Similarly, as
a result of Stone-Wales type transformation, each C$_{4}$
defect region also turns into a nonagon-shaped hole. As it
was reported experimentally,\cite{hashimoto} a hole region is
surrounded by pentagonal and hexagonal rings of C atoms to
keep the flatness of the sheet. While the honeycomb lattice
symmetry of the graphene matrix does not change considerably
for larger defects C$_{6}$ and C$_{24}$, hole region of
C$_{12}$
takes almost a circular shape surrounded by regular pentagons
and hexagons. Apparently, the trend in the shape of the hole
region is determined by whether the edges of domain are
zigzag or armchair shaped.

\begin{figure}
\includegraphics[width=8.5cm]{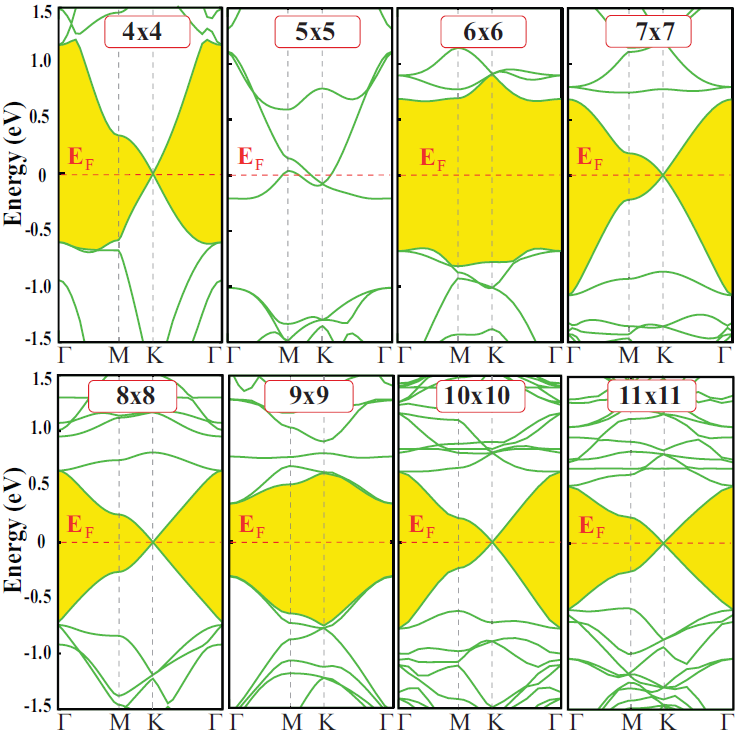}
\caption{ Band structure of nanomeshes of C$_{12}$ forming  in the
($n$x$n$) supercells of graphene with $n$=4...11.}
\label{f5}
\end{figure}

The energy band structures of graphene nanomeshes exhibit
interesting variations with respect to their sizes, diameters
and shapes of nanohole as illustrated in Fig.\ref{f4} (b).
While nanomesh of C$_{1}$ defect have metallic behavior,
periodic structure of C$_{2}$ defect becomes indirect band gap
semiconductor of 0.65 eV. However, $\pi$ and $\pi^{*}$-bands
above the Fermi level still very close to each other (0.09 eV
gap) at the $K$-point. C$_{4}$ defected GNM has metallic
behavior. The situation becomes even more remarkable for
larger defects C$_{6}$, C$_{12}$ and C$_{24}$. GNMs
including either C$_{6}$ or C$_{24}$ hole have zigzag edges
and are metals with antiferromagnetic (AFM) ground state. We note that the defect induced flat electronic bands around the Fermi level occur if the electron spins become  unpaired. On the other hand, the reconstruction or the dimerization of carbon dangling bonds around the defect can cause the flat bands in the gap to disappear. Interestingly, nanomeshes of C$_6$ and C$_{24}$ holes, which are metallic, become semimetal upon the saturation of dangling bonds of carbon
atoms around the hole by hydrogen atoms. This metal-semimetal
transformation can have applications in graphene-based nanoelectronics.
On the other hand, GNM of C$_6$ turns semiconducting with a band gap of
0.1 eV upon the termination of the dangling bonds of C atoms
around the hole by B and N alternatingly to from a B$_3$N$_3$
hexagon. The band opening is explained by the breaking of
sixfold symmetry due to B$_3$N$_3$ hexagon. The formation extended B$_n$N$_n$ 
honeycomb structure can be achieved directly in the course of epitaxial growth 
of graphene and single layer BN.\cite{seymurCS, CS2}

As for GNM with C$_{12}$, it is a nonmagnetic semimetal because of
carbon atoms at the edge are dimerized. The analysis of the orbital
character of linearly crossing $\pi$- and $\pi^{*}$-bands near
$K$-point using isosurface charge densities suggests that these
bands originate from bonding and antibonding combinations of $\pi$-
orbitals at the neck and around C$_6$ hole. Here we discuss an
important aspect of GNMs with C$_{12}$, that the size of the hexagonal supercell
or repeat periodicity of C$_{12}$ is crucial for the resulting electronic
structure. Fig. \ref{f5} shows band structures of GNMs including
single C$_{12}$ hole in the supercell of ($n$x$n$) where $n$=4...11.
For $n$=4, GNM has the neck region consisting of single hexagon
is a semimetal. For $n$=5, GNM is a metal and has a neck region
which is relatively ticker, but its C-C bond angles strongly
deviate from 120$^o$. Surprisingly, GNM with $n$=6 is a semiconductor
having 1.3 eV band gap. The bond angles still continue to deviate
from those of graphene. However, from $n$=7 on the bond angles at
the neck region start to be graphene like with regular honeycomb
structure. Both GNMs with $n$=7 and 8 are semimetals and  have
$\pi$- and $\pi^*$-bands which linearly cross at Fermi level
at the $K$-point.  Isosurface charge densities of these bands near
$K$-points demonstrate that they, in fact, originate from the combination of
graphene $\pi$-orbitals. As $n$ increases, GNM exhibit the similar
trend for $n$=6-9; namely it is semiconductor for $n$=9, but semimetals
for $n$=10 and 11. This variation of band gaps is reminiscent of the
family behavior of graphene nanoribbons and is related to the variation
of the thickness of necks between periodically repeating C$_{12}$ holes.\cite{topsakal, seymur}
Here, even if the six fold rotation symmetry is conserved, the band gap
opens for every $n$=3x$N$ with $N$ being an integer $\geq 2$. However, this
gap becomes smaller and eventually is closed as $n \rightarrow \infty$.
We also note that the family like behavior of GNMs is related with the
edge structure of the hole. In regard to the size of GNM of C$_{12}$, we
note also that the lattice constants of corresponding  ($n$x$n$) supercell
is modified with size. For example, for $n$=11, the lattice constant of
supercell is contracted by 1\%, the contraction is 4\% for $n$=5 and 40\%
for $n$=4.

\begin{figure}
\includegraphics[width=8.5cm]{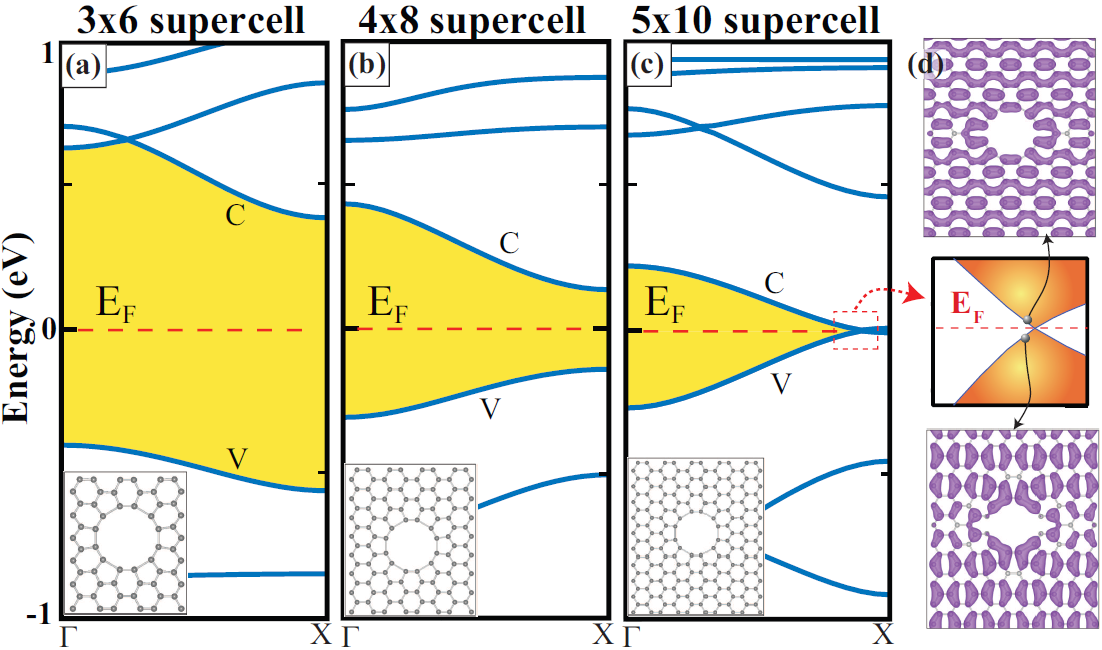}
\caption{(a)-(c) Rectangular patterns of C$_{12}$ holes repeated in (3x6), (4x8) and (5x10) graphene supercells. Atomic structure of nanomeshes are given by inset. (d) Semimetallic electronic structure and isosurface charge densities of valence (V) and conduction (C) bands of (5x10) nanomesh are also shown.}
\label{f6}
\end{figure}

Finally, in addition to triangular defect patterns, we also
discuss the electronic structure of holes arranged in a
rectangular lattice. In Fig. \ref{f6} we show the electronic
band structures of C$_{12}$ nanomeshes realized by the
supercells of (3x6), (4x8) and (5x10). While C$_{12}$ holes
in a small rectangular supercells with small repeat
periodicity (leading to significant coupling) become
semiconducting, the semimetallic nature indigenous to
graphene is achieved in large supercells. As shown in Fig.
\ref{f6}(c), even if the rotation symmetry required for band
crossing is absent, graphene-like Bloch wave functions in the
rectangular mesh of sparse patterns of C$_{12}$ holes show a
semimetallic behavior.  This indicates that as the size of
the supercell becomes larger and the neck gets wider relative
to the size of the hole, the symmetry requirement necessary
for the linear band crossing can be relaxed.

\section{Mechanical Properties of Nanomeshes}

Honeycomb structure with $sp^{2}$ bonding underlies the
unusual mechanical properties providing very high in-plane
strength, but transversal flexibility. Here we investigate
how the mechanical properties of nanomeshes generated with
patterns of adatoms or holes. We focused on the harmonic
range of the elastic deformation, where the structure
responded to strain $\epsilon$ linearly. Here $\epsilon$
the elongation per unit length. The strain energy is defined
as $E_{s}=E_{T}(\epsilon)-E_{T}(\epsilon=0)$; namely, the
total energy at a given strain $\epsilon$ minus the total
energy at zero strain. Normally, the Young's modulus is the
value, which characterizes the mechanical strength of a bulk
material. Owing to ambiguities in defining the Young's
modulus of two dimensional structures like a GNM, one can use
in-plane stiffness $C=(1/A_{0})\cdot(\partial^{2}E_{S}/\partial\epsilon^{2})$ in
terms of the equilibrium area of the supercell, $A_{0}$.\cite{yakobson1,reddy} We calculated
the in-plane stiffness of graphene, and nanomeshes consisting
of C$_{32}$H$_6$ , BN substituted graphene (i.e C$_{26}$[(BN)$_3$] and
C$_6$ hole in graphene using (4x4) supercell in Fig. \ref{f3}. The stiffness
of bare graphene is calculated to be 334 N/m, which is in good
agreement with the experimental value of 340$\pm$50 N/m.
Furthermore, the in-plane stiffness values of nanomeshes
generated on graphene through B$_3$N$_3$ substitution, H$_6$
adatom pattern and C$_6$ hole are 308 N/m, 283 N/m and 167
N/m, respectively. Apparently, the bare graphene matrix is
weakened by the formation of any of these nanomeshes. In
addition to the calculation of in-plane stiffness, we extend
our analysis to include the plastic deformation region, where
the honeycomb like structure is destroyed after the yielding
point (i.e. on set of plastic deformation), and GNM undergoes
a massive structural deformation. Our preliminary simulations
indicate that the yielding strain of C$_6$ hole GNM is significantly
lower than the yielding strains of both C$_{26}$[(BN)$_3$] and
C$_{32}$H$_6$ GNMs.

\section{Conclusions}

While graphene and its various derivatives, GO, CH and CF, are important nanomaterials with diverse electronic, magnetic and mechanical properties, their properties can be modified and multiplied using different methods of functionalization. The most pronouncing property of graphene, namely linearly crossing bands at the Fermi level and electron-hole symmetry arising therefrom is usually destroyed, when graphene is functionalized through dopant or vacancy defects. In this work we demonstrated that the electron-hole symmetry, even Dirac Fermion behavior can be recovered for periodically repeating superstructures or nanomeshes having special point group symmetry. In this study we considered nanomeshes, which are generated by the decoration of adatoms, adatom groups or holes, which repeats periodically in graphene matrix. We found that types of adatoms and their patterns, the geometry of the holes of carbon atoms, the sizes and lattice symmetries of nanomesh provide us with several parameters to engineer the electronic and magnetic properties of the nanomesh. In particular, we showed that by varying only the size of the nanomesh including a specific hole one can tune between metallic and semiconducting state including semimetal with linearly crossing bands. This is reminiscent of the family behavior of graphene nanoribbons.

\section{ACKNOWLEDGMENTS}

This work is supported by TUBITAK through Grant No:108T234.
Part of the computational resources has been provided by TUBITAK ULAKBIM, High Performance and Grid Computing Center (TR-Grid e-Infrastructure). We also thank the DEISA Consortium (www.deisa.eu), funded through the EU FP7
project RI-222919, for support within the DEISA Extreme Computing Initiative. S. C. acknowledges the partial support of TUBA, Academy of Science of Turkey.

\end{document}